\documentstyle[preprint,eqsecnum,aps]{revtex}
\begin{document}
\draft
\title{ Statistics for Particles Having Internal Quantum State }
\author{ Zhi-Tao Yan\footnote{E-mail: yanzt@ibm320h.phy.pku.edu.cn} }
\address{  Department of Physics, Peking University,
	 Beijing 100871, P. R. China   }
\date{November 9, 1997}
\maketitle

\begin{abstract}
A new kind of quantum statistics which interpolates between Bose and Fermi
statistics is proposed beginning with the assumption that
the quantum state of a many-particle system is a functional on the
internal space of the particles.
The quantum commutation relations for such particle creation and
annihilation operators are derived, and statistical partition function
and thermodynamical properties of an ideal gas of the particles are
investigated. The application of this quantum statistics for the ensemble
of extremal black holes are discussed.
\end{abstract}


Although there were some attempts to propose a generalized quantum
statistics (GQS)~\cite{Gentile}, yet Bose and Fermi statistics were believed
to be unique for quite a long time. However, since the model by Wilczek~\cite
{Wilczek} which is system of particles with an Aharonov-Bohm type of
interaction in two dimensions, particles called anyons have been a subject
of intense study and a number of different physical applications have been
investigated, such as fractional quantum Hall effect (FQHE)~\cite{Halperin}
and high-tempreture superconductivity~\cite{Laughlin}. The concept of
anyons, which is based on the wave function arises a factor e$^{i\phi }$ as
it exchanges two particles (exchange statistics), is essentially
two-dimensional. Another way to define GQS has been formulated by Haldane%
\cite{Haldane}, which is based on the rate at the number of the available
states in a system of fixed size decrease as more and more particles add to
it (exclusion statistics). This statistics, formulated without any reference
to spatial dimensions, captures the essential features of the anyon
statistics peculiar to two-dimensional systems. Recently, a variant notion
of GQS bases on deformations of the bilinear Bose and Fermi commutation
relations. Particles obeying a simple case of this type statistics (the
so-called ''infinite'' statistics) are called quons\cite{Greenberg}, which
obey the minimally deformed commutator 
\begin{equation}
\lbrack a_i,a_j^{\dagger }]_q=\delta _{ij}  \label{quon}
\end{equation}
where $[A,B]_q\equiv AB-qBA$, and $q$ is a $c$-number, $|q|\le 1$. The
equivalence of anyon statistics and quon statistics, Eq.(\ref{quon}) with $%
q=e^{i\phi }$, was proved~\cite{Goldin} via the properties of the $N$-anyon
permutation group. More recently, a new model of GQS, in which identical
particles exhibit both Bose and Fermi statistics with respective
probabilities $p_b$ and $p_f$, is introduced by Medvedev\cite{Medvedev}.

In this letter we investigate a new exchange statistics beginning with the
assumption that the quantum state of a many-particle system is a functional
on the internal space of the particles. Three decades ago, it was shown by
Finkelstein and Rubenstein~\cite{Fin} that, in nonlinear field theories,
soliton statistics can be determined from the fact that the quantum state is
a functional on the space of field configurations. If the soliton has no
internal states, the eigenspaces of exchange operator are superselection
sector: Bosons are forever bosons and fermions are forever fermions; if the
soliton does have internal state then the exchange operator may or may not
change the field configuration, depending on whether or not the solitons are
the same state\cite{CTQS}. These ideas were firstly applied to quantum
gravity in a series of beautiful papers by Friedman and Sorkin\cite{Friedman}%
. Recently, Strominger applied these ideas to the problem of charged
extremal black hole statistics and he argued that the charged extremal black
holes will obey the infinite statistics with $q=0$ on the condition that
none of them are in the same internal state.

Basing on the developed ideas above, we assume at first that the wave
function of such particles is more composition, {\it a functional}, according
to its internal degrees of freedom, and suppose then that a {\it phase factor%
} the wave functional of the many-particle system arises as single exchange
of a pair of particles is an {\it operator}~\cite{Sci}, which is dependent
on the intrinsic property of the particles, rather than an usual $c$-number.
Furthermore, we consider that all the particles of the system are in the
same state, and then the exchange operator will be independent of the pairs
of particles and commute with any operator in the system simply because it
dose neither change the physical configuration nor mix up internal states
again. This operator is marked by $\hat q$ in following.

Consider a system of $N$ noninteracting such identical particles,
represented by wave functional $\Psi (x_1,\cdots ,x_j,\cdots ,x_i,\cdots
,x_N)$. Single exchange any two particles we get 
\begin{equation}
\Psi (x_1,\cdots ,x_j,\cdots ,x_i,\cdots ,x_N)=\hat q\Psi (x_1,\cdots
,x_i,\cdots ,x_j,\cdots ,x_N)  \label{UHsymm}
\end{equation}
It is easy to show that the operator $\hat q$ is both Hermitian and unitary~%
\cite{WuTY}, i.e., it satisfies 
\begin{equation}
\hat q=\hat q^{\dagger }\ \ \ {\rm and}\ \ \ \ \hat q^{\dagger }\;\hat
q=\hat q\;\hat q^{\dagger }=\;{\bf 1}
\end{equation}
The eigenvalues of $\hat q$ take on $+1$ or $-1$ only, and the eigenequation
of $\hat q$ is 
\begin{equation}
\hat q\ |\pm 1,j>\ =\ \pm 1\ |\pm 1,j>
\end{equation}
where $|\pm 1,j>$ compose a complete orthonormal set, with $j$ denoting the
degeneration degrees of freedom; and then the internal quantum state can be
written 
\begin{equation}
|{\cal I}n>=\sum_ic_i^{(+)}|+1,i>+\sum_jc_j^{(-)}|-1,j>  \label{giqs}
\end{equation}
with the normalization condition: $\sum_i|c_i^{(+)}|^2+\sum_j|c_j^{(-)}|^2=1$%
. The exchange symmetry in Eq.(\ref{UHsymm}) is generally neither symmetry
nor anti-symmetry and we call it $\hat q$-symmetry below.

To study statistical properties of the system, let us first develop the
notation of the occupation-number representation. For convenience, we
introduce the following Hermitian projection operator for the system 
\begin{equation}
\hat Q=\frac 1{N!}\sum_P\hat q^{[P]}P
\end{equation}
where $P$ is permutation operator, the summation is over the $N!$
permutations of the $N$ particles and $\hat q^{[P]}={\bf 1}$ or $\hat q$
according as the permutation $P$ is {\it even {\rm or} odd}. It is easily
proved \cite{WuTY} that 
\begin{equation}
\hat Q^{\dagger }=\hat Q\;\;\;\;\hat Q^2=\hat Q  \label{Q}
\end{equation}

Let $\{ \psi_i \}$ be a complete orthonormal set of single-particle states,
the $\hat{q}$-symmetrized wave function can be expressed in the form 
\begin{equation}  \label{Qsw}
\Psi^{\hat{q}}(x_1,x_2,\cdots ,x_N) = C_N^{\hat{q}} \hat{Q} \{
\psi_{i_1}(x_1) \psi_{i_2}(x_2) \cdots \psi_{i_N}(x_N) \}
\end{equation}
where $C_N^{\hat{q}}$ is the {\it normalization constant} to be specified
below. If there are $n_1$ particles in single-particle state $\psi_1$, $n_2$
particles in single-particles state $\psi_2$, etc., then not all of $N!$
terms $\hat{Q}\{ \psi_{i_1}(x_1) \psi_{i_2}(x_2) \cdots \psi_{i_N}(x_N) \}$
are, in general, different. Take the permutations only among $\psi_i$ (given
$n_i>1$) single-particle states as an example, we get $\frac{n_i!}{2}$
permutations are in the same form and the other $\frac{n_i!}{2}$
permutations are in another same form which is only different from the form
above by a factor $\hat{q}$. Consequently, if the internal permutations of $%
\psi_i$s are not considered, one can equivalently contribute a factor $\frac{%
n_i!}{2}({\bf 1}+\hat{q})$. Thus the sum of all $N!$ permutations are equal
to the sum of all the {\it different} permutations products a factor $%
(\prod_in_i!)(\frac{{\bf 1}+\hat{q}}{2})^T$, where $T$ is the total number
of single-particle state occupied by more than one particle. Hence Eq.(\ref
{Qsw}) becomes 
\begin{eqnarray}  \label{Psiq}
\Psi^{\hat{q}}_{n_1,n_2,\cdots,n_k}(x_1,x_2,\cdots,x_N) & = & C_N^{\hat{q}}
(\prod_{i=1}^kn_i!)(\frac{{\bf 1}+\hat{q}}{2})^{T} \frac{1}{N!}\sum_{P_E}{%
\hat{q}}^{[P_E]} P_E\{ [\psi_1(x_1)\cdots \psi_1(x_{n_1})]  \nonumber \\
& & [ \psi_2(x_{n_1+1}) \cdots \psi_2(x_{n_1+n_2}) ] \cdots [ \psi_k \cdots
\psi_k(x_N) ] \}
\end{eqnarray}
where $P_E$ are the permutations only among {\it different} single-particle
states, $\hat{q}^{[P_E]} ={\bf 1}$ or $\hat{q}$ according as the permutation 
$P_E$ is {\it even {\rm or} odd} and all these permutation terms are
different.

Now we introduce the unit operator, marked by ${\bf 1}^{\hat{q}}$, for the
system, and then $C_N^{\hat{q}}$ can be factorized as $C_N^{\hat{q}}=C_N\;%
{\bf 1}^{\hat{q}}$. From Eqs.(\ref{Q}), (\ref{Qsw}) and (\ref{Psiq}), we
obtain
\begin{equation}
{\bf 1}^{\hat{q}}=(\Psi^{\hat{q}}, \Psi^{\hat{q}})=|C_N|^2 {\bf 1}^{\hat{q}} 
\frac{\prod_{i=1}^{k} n_i!}{N!} (\frac{{\bf 1}+\hat{q}}{2})^{T}
\end{equation}
By using Eq.(\ref{UHsymm}), one then obtain from this equation that 
\begin{equation}  \label{1q}
{\bf 1}^{\hat{q}}=\left \{ 
\begin{array}{ll}
{\bf 1} & {\rm for } \; n_1,n_2,\cdots,n_i,\cdots\le 1 \\ 
\frac{1}{2}({\bf 1}+\hat{q}) & {\rm otherwise\; (i.e.,exist}\; n_i \ge 2)
\end{array}
\right.
\end{equation}
and $C_N= \sqrt{\frac{N!}{\prod_{i=1}^k n_i!}} $. Thus 
\begin{eqnarray}  \label{qsos}
\Psi^{\hat{q}}_{n_1,n_2,\cdots,n_k}(x_1,x_2,\cdots,x_N) & = & \sqrt{\frac{%
\prod_{i=1}^k n_{i}!}{N!}}\; {\bf 1}^{\hat{q}}\;\sum_{P_E} \hat{q}^{[P_E]}
P_E\{ [\psi_1(x_1)\cdots \psi_1(x_{n_1})]  \nonumber \\
& & [ \psi_2(x_{n_1+1}) \cdots \psi_2(x_{n_1+n_2}) ] \cdots [ \psi_k \cdots
\psi_k(x_N) ] \}
\end{eqnarray}
is a $\hat{q}$-symmetrized orthonormal $N$-particle state.

In the occupation-number representation the state Eq.(\ref{qsos}) is written 
$|n_1,n_2,\cdots,n_k>^{\hat{q}}$, where $n_i$ is the number of particles in
the state $\psi_i$ and the superscript $\hat{q}$ marks intrinsic property of
the particle (for instance, when $\hat{q}=$ {\bf 1} or {\bf -1}, the
particle is bosonic or fermionic). Subject to $\sum_in_i=N$, all of $%
|n_1,n_2,\cdots>^{\hat{q}}$ form a complete orthonormal set, i.e., 
\begin{equation}
\ ^{\hat{q}}<n_1^{\prime},n_2^{\prime},\cdots|n_1,n_2,\cdots>^{\hat{q}}= 
{\bf 1}^{\hat{q}} \delta_{n_1n_1^{\prime}}\delta_{n_2n_2^{\prime}}\cdots
\end{equation}
\begin{equation}  \label{comset}
\sum_{ \{n_i\},\;\sum_in_i=N }|n_1,n_2,\cdots>^{\hat{q}} \ ^{\hat{q}%
}<n_1,n_2,\cdots|={\bf 1}^{\hat{q}}
\end{equation}
and compose the basis states of a Hilbert space ${\cal H}^{\hat{q}}_N$.

To simplify application of the grand-canonical ensemble formulation of
statistical mechanics, we extend the development above to systems with an
indefinite number of particles, whose corresponding Hilbert space is called
Fock-like space here. It is useful to define the vacuum state, denoted $%
|0>^{\hat q}$, which represents a state $|0,0,\cdots ,0,\cdots >^{\hat q}$
with zero particles. 
Hence, the Fock-like space can be indicated: ${\cal F}^{\hat
q}=\bigoplus_{N=0}^{N=\infty }{\cal H}_N^{\hat q}$, where by definition: $%
{\cal H}_0^{\hat q}=|0>$. The closure relation in the Fock-like space may be
written 
\begin{equation}
\sum_{n_1,n_2,\cdots ,n_i,\cdots }|n_1,n_2,\cdots ,n_i,\cdots >^{\hat q}\
^{\hat q}<n_1,n_2,\cdots ,n_i,\cdots |={\bf 1}^{\hat q}  \label{clos}
\end{equation}
We now introduce the annihilation operator $\hat a_i$ and its Hermitian
conjugate creation operator $\hat a_i^{\dagger }$ in the Fock-like space.
They are defined in term of their effects on the state vector, as follows: 
\begin{equation}
\hat a_i|n_1,n_2,\cdots ,n_i,\cdots >^{\hat q}=\hat q^{(\sum_{l=1}^{i-1}n_l)}%
\sqrt{n_i}{\hat q}\;{\bf 1}^{\hat q}|n_1,n_2,\cdots ,n_i-1,\cdots >^{\hat q}
\label{qann}
\end{equation}
and 
\begin{equation}
\hat a_i^{\dagger }|n_1,n_2,\cdots ,n_i,\cdots >^{\hat q}=\hat
q^{(\sum_{l=1}^{i-1}n_l)}\sqrt{n_i+1}\;{\bf 1}^{\hat q}|n_1,n_2,\cdots
,n_i+1,\cdots >^{\hat q}  \label{qcre}
\end{equation}
where the associated factor $\hat q^{(\sum_{l=1}^{i-1}n_l)}$ comes from the
exchange symmetry Eq.(\ref{UHsymm}) among different single-particle states
and the unit operator ${\bf 1}^{\hat q}$ is written distinctly in order that
the following calculation becomes clear. By using Eq.(\ref{qcre}), it is easy
to show that the Fock-like space ${\cal F}^{\hat q}$ also can be spanned by
$|0>^{\hat q}$ and all $\frac{(\hat a_1^{\dagger})^{n_1}} {\sqrt{n_1!}}%
\frac{(\hat a_2^{\dagger })^{n_2}}{\sqrt{n_2!}}\cdots |0>^{\hat q}$.
We next define the number operator for particles in the
state $\psi _i$ by $\hat n_i=\hat a_i^{\dagger }\hat a_i$. Then, one can
easily find that 
\begin{equation}
\hat n_i|n_1,n_2,\cdots ,n_i,\cdots >^{\hat q}=n_i{\bf 1}^{\hat
q}|n_1,n_2,\cdots ,n_i,\cdots >^{\hat q}
\end{equation}
The total number operator is clearly $N(\hat q)=\sum_i\hat n_i=\sum_i\hat
a_i^{\dagger }\hat a_i$. A careful application of Eqs.(\ref{qcre}), (\ref
{qann}) and (\ref{1q}) yields 
\begin{equation}
\lbrack \hat a_i,\hat a_j^{\dagger }]_{\hat q}={\bf 1}^{\hat q}\delta _{ij}
\label{qc}
\end{equation}
\begin{equation}
\lbrack \hat a_{i},\hat a_{j}]_{\hat q}=[\hat a_{i}^{\dagger },\hat
a_{j}^{\dagger }]_{\hat q}=0
\end{equation}
The commutators are similar to the so-called $\hat q$-uon commutators
introduced by Wu and Sun~\cite{WuLA} in the point that the $c$-number $q$
in Eq.(\ref{quon}) is replaced by a linear
operator $\hat{q}$, but the real difference
between them is that in Eq.(\ref{qc}) the creation and annihilation operator
and their commutator resulting from the unit operator Eq.(\ref{1q}) are
number-distribution-dependent. It is easily shown that the commutators are
peculiar to the commutators of Boson when $\hat q={\bf 1}$ and Fermion when $%
\hat q={\bf -1}$.

Following, we investigate the statistical properties of such particles.
Naturally, the Hamiltonian of an ideal gas of such particles is given by $%
H(\hat q)=\sum_i\varepsilon _i\hat a_i^{\dagger }\hat a_1$, with $%
\varepsilon _i$s the single-particle energy levels. Now let us start with
the grand canonical partition functional 
\begin{equation}
\Xi (\hat q)={\rm Tr}_{\hat q}\ \exp \{-\beta [H(\hat q)-\mu N(\hat q)]\}
\end{equation}
where $\beta =1/kT$ with $T$ being temperature of the gas, $\mu $ is the
chemical potential, and ${\rm Tr}_{\hat q}\ (...)=\sum_{n_1,n_2,\cdots }\
^{\hat q}<n_1,n_2,\cdots |\ (...)\ |n_1,n_2,\cdots >^{\hat q}$. From $%
e^{\hat A}=\sum_{n=0}^\infty \frac{\hat A^n}{n!}$ and Eq.(\ref{UHsymm}),
the partition functional can be derived
\begin{equation}
\Xi (\hat q)=\prod_i\frac{1-\frac{1-\hat q}2e^{2\beta (\mu -\varepsilon _i)}%
}{1-e^{\beta (\mu -\varepsilon _i)}}
\end{equation}

It is evident that, besides the intrinsic operator $\hat q$ mentioned above,
the physical properties of such particles are decided by internal quantum
state which reflects the statistics of identical particles with internal
degrees of freedom. For instance, when the internal state of the particle is
the eigenstate of $\hat q$,\ $\sum_ic_i^{(+)}|+1,i>$ or $%
\sum_jc_j^{(-)}|-1,j>$, the particle displays boson or fermion. In general,
the internal quantum state of such particles is Eq.(\ref{giqs}), and then the
particles show a new feature which interpolates between boson and fermion.
Under such state, we obtain the grand partition function 
\begin{equation}
\Xi (\delta )=\prod_i\frac{1-\delta z^2e^{-2\beta \varepsilon _i}}{%
1-ze^{-\beta \varepsilon _i}}
\end{equation}
where $\delta =\sum_j|c_j^{(-)}|^2$,\ $0\le \delta \le 1$ and $z=e^{\beta
\mu }$. The meaning of the parameter $\delta $ is similar to the probability 
$p_f$ in Medvedev's model~\cite{Medvedev}. Obviously, when $\delta =0$ or $1$%
, we obtain the partition function of ideal Boson or Fermion~\cite{Pathria}.
The occupation number in energy level $\varepsilon _i$ is
\begin{equation}
n_i=\frac 1{e^{\beta (\varepsilon _i-\mu )}-1}-\frac{2\delta }{e^{2\beta
(\varepsilon _i-\mu )}-\delta }
\end{equation}
In order for all possible $n_i$ is non-negative, the chemical potential
should be satisfy 
\begin{equation}
\mu \le 0\;\;\;\;{\rm or}\;\;\mu \ge \;\varepsilon _{max}-\frac 12kT\ln
\delta  \label{mu}
\end{equation}
where $\varepsilon _{max}$ is the highest energy level of the system. Thus
the chemical potential of this ideal gas will emerge a gap on condition that 
$\varepsilon _{max}-\frac 12kT\ln \delta >0$.

The thermodynamic properties can be derived straight-forwardly from
partition function. The thermodynamic potential, $\Omega =-PV$, is given by
\begin{equation}
\Omega =-\beta ^{-1}\ln \Xi (\delta )=-\beta ^{-1}\sum_i[-\ln (1-z{\rm e}%
^{-\beta \epsilon _i})+\ln (1-\delta z^2{\rm e}^{-2\beta \epsilon _i})]
\label{PV}
\end{equation}
The total particle number of the system is 
\begin{equation}
N=\sum_i\ [\frac 1{e^{\beta (\varepsilon _i-\mu )}-1}-\frac{2\delta }{%
e^{2\beta (\varepsilon _i-\mu )}-\delta }]  \label{N}
\end{equation}
The chemical potential $\mu $ as a function of temperature $T$ and number
density $\frac NV$ is defined implicitly by Eqs.(\ref{mu}) and (\ref{N}).
Since the summation is converted to integration as $\sum_{{\bf p}%
}\rightarrow \frac V{h^3}\int d^3p$, if one assumes such free particles $%
\varepsilon _{{\bf p}}=\frac{{\bf p}^2}{2m}$, $m$ is the mass of the
particle, then the density of state is given by $a(\varepsilon )=\frac
V{h^3}2\pi (2m)^{3/2}\varepsilon ^{1/2}$ and $\sum_{{\bf p}}\rightarrow \int
d\varepsilon a(\varepsilon )$. After some mathematical manipulation~\cite
{Pathria}, we get 
\begin{equation}
\frac{PV}{kT}=\frac V{\lambda ^3}[g_{5/2}(z)-\frac
1{2^{3/2}}g_{5/2}(z^{\prime })]-\ln (\frac{1-z^{\prime }}{1-z})
\end{equation}
and 
\begin{equation}
N=\frac V{\lambda ^3}[g_{3/2}(z)-\frac 1{2^{1/2}}g_{3/2}(z^{\prime
})]+(\frac z{1-z}-\frac{2z^{\prime }}{1-z^{\prime }})
\end{equation}
where $z^{\prime }=\delta z^2$ , $\lambda =\frac h{{(2\pi mkT)}^{1/2}}$, and 
$g_n(z)\equiv \frac 1{\Gamma (n)}\int_0^\infty \frac{x^{n-1}dx}{z^{-1}e^x-1}$
is the Bose-Einstein integrals. In the classical Boltzmann limit $z<<1$ the
equation of the state taked the form of the {\it virial expansion}\ to the
third order is 
\begin{equation}
PV=NkT\{1+\frac 12(\frac \delta {\sqrt{2}}-\frac 1{2\sqrt{2}})\frac{Nh^3}{%
V(2\pi mkT)^{3/2}}+[(\frac \delta {\sqrt{2}}-\frac 1{2\sqrt{2}})^2-\frac 2{9%
\sqrt{3}}]\frac{N^2h^6}{V^2(2\pi mkT)^3}\}  \label{stateeq}
\end{equation}
There is an effective attraction between such ideal gas when\ $\delta <1/2$
and effective repulsion when$\ \delta >1/2$. In the case $\delta =1/2$, a
weak attraction still exists due to the third term.

As was mentioned above, in the case that none of them are in the same
internal state (where the internal state-number must be larger than the
particle-number), charged extremal black holes will obey the infinite
statistics with the deformation parameter $q=0$\cite{Strominger}.
Correspondingly, in the case that all of them are in the same internal
state(with no restriction on internal state-number), the black holes will
obey the new kind of statistics. Consider a collection of charged,
nonrotating extremal black holes which have the same internal state. In the
dilutal nonrelativistic gas approximation, if gravitational and
electrostatic interactions reach equilibrium via Hawking radiation, then the
system is a gas of {\it noninteracting neutral particles}. So its
thermodynamical properties are governed by the particle statistical
properties, alone. According to Eq.(\ref{stateeq}), if the internal state
of the charged extremal black holes satisfy $\delta \le 1/2$, then they will
experience {\it weak statistical attraction }. Hence, one may also expect
that the black holes will form clusters as expected by
Medvedev~\cite{Medvedev}.

In summary, we have introduced a new quantum statistics, which interpolates
between Bose and Fermi statistics smoothly, via the assumption that all of
the particles are in the same internal quantum state reflecting the
statistics of identical particles with internal degrees of freedom. A
Fock-like Space of this system is obtained and the quantum commutation
relations for such particle creation and annihilation operators are derived.
Partition function and thermodynamical properties of an ideal gas of the
particles are investigated. Finally, we discussed the implication of this
statistics to the collection of charged extremal black holes having
the same internal quantum state.

I would like to thank Professor L.-A. Wu for some useful comments and
bringing Ref\cite{WuLA} to my attention, and I also thank Professor C.-S.
Gao and X.-C. Xu for useful help and discussions.

\end{document}